\documentclass[conference,a4paper]{IEEEtran}

\usepackage{cite}
\usepackage{graphicx,color,epsfig,rotating}
\usepackage{amsfonts,amsmath,amssymb,bbm}
\usepackage{algorithm, algorithmic}
\usepackage{subfigure}
\usepackage{amsmath}
\usepackage{cite}
\usepackage{mdwtab}
\usepackage{subfigure}
\usepackage{placeins}
\usepackage{psfrag, graphicx}
\usepackage[latin1]{inputenc}
\usepackage{amssymb}
\usepackage{makeidx}
\usepackage{epstopdf}

\long\def\comment#1{}

\newfont{\bbb}{msbm10 scaled 700}

\newfont{\bb}{msbm10 scaled 1100}

\newcommand{\be}{\begin{equation}}
\newcommand{\ee}{\end{equation}}
\newcommand{\bea}{\begin{eqnarray}}
\newcommand{\eea}{\end{eqnarray}}

\newtheorem{theorem}{Theorem}
\newtheorem{remark}{Remark}

\begin{document}

\title{Coded Distributed Computing with Heterogeneous Function Assignments}

\author{Nicholas Woolsey, Rong-Rong Chen and Mingyue Ji
\thanks{The authors are with the Department of Electrical Engineering,
University of Utah, Salt Lake City, UT 84112, USA. (e-mail: nicholas.woolsey@utah.edu, rchen@ece.utah.edu and mingyue.ji@utah.edu)}
}

\author{
    \IEEEauthorblockN{ Nicholas Woolsey,
		Rong-Rong Chen, and Mingyue Ji }
	\IEEEauthorblockA{Department of Electrical and Computer Engineering, University of Utah\\
		Salt Lake City, UT, USA\\
		Email: \{nicholas.woolsey@utah.edu,
		 rchen@ece.utah.edu,
		mingyue.ji@utah.edu\}}

}

\maketitle

\thispagestyle{empty}
\pagestyle{empty}

\vspace{-0.5cm}

\begin{abstract}
Coded distributed computing (CDC) introduced by Li \textit{et. al.} is an effective technique to trade computation load for communication load in a MapReduce framework. CDC achieves an optimal trade-off by duplicating map computations at $r$ computing nodes to yield multicasting opportunities such that $r$ nodes are served simultaneously in the Shuffle phase. However, in general, the state-of-the-art CDC scheme is mainly designed only for homogeneous networks, where the computing nodes are assumed to have the same storage, computation and communication capabilities. In this work, we explore two novel approaches of heterogeneous CDC design. First, we study CDC schemes which operate on multiple, collaborating homogeneous computing networks. Second, we allow heterogeneous function assignment in the CDC design, where nodes are assigned a varying number of reduce functions. Finally, we propose an expandable heterogeneous CDC scheme where $r-1$ nodes are served simultaneously in the Shuffle phase. In comparison to the state-of-the-art homogeneous CDC scheme with an equivalent computation load, we find our newly proposed heterogeneous CDC scheme has a smaller communication load in some cases.
\end{abstract}



\section{Introduction}
\label{section: intro}

Coded distributed computing (CDC), introduced in \cite{li2018fundamental}, offers an efficient approach to reduce the communication load in  CDC networks such as MapReduce \cite{dean2008mapreduce}. 
In this setting, $K$ computing nodes are assigned to compute $Q$ functions, where each function requires $N$ files as input. In general, each computing node does not have access to all $N$ files, and therefore, computation is split into ``Map" and ``Reduce" phases. In the Map phase, using map functions, nodes compute intermediate values from their locally available files. Then, in the Reduce phase, the intermediate values are used to compute reduce functions to obtain the desired function outputs. As nodes require intermediate values that cannot be computed locally, the nodes transmit intermediate values amongst one another in the ``Shuffle" phase which occurs in between the Map and Reduce phases. Often times, the Shuffle phase takes up a majority of the overall MapReduce execution time. To alleviate this bottleneck using the state-of-the-art CDC scheme, map computations are repeated at $r$ carefully chosen nodes to reduce the communication load by a factor of $r$ \cite{li2018fundamental}.


There are $3$ important design considerations while developing a CDC scheme which include: file assignment, reduce function assignment and Shuffle phase design. 
For example, in the state-of-the-art scheme \cite{li2018fundamental}, the $N$ files are split into ${K \choose r}$ disjoint, equal-size file sets and each set is assigned to a unique set of $r$ nodes. Furthermore, the reduce functions are split into $K$ disjoint, equal-size subsets and each node is assigned one of the function sets. This specific file and function assignment creates multicasting opportunities where a single transmission can simultaneously serve $r$ nodes in the Shuffle phase. While this scheme, and other CDC schemes \cite{konstantinidis2018leveraging,woolsey2018new}, obtain an optimal, or near optimal, computation-communication load trade-off, their scope is limited by an underlying assumption that the computing network is homogeneous. In other words, each node is assigned the same number of files and functions and considered to have the same storage and computation capabilities. However, in general, computing networks are often heterogeneous in nature.


Designing a CDC scheme that fully utilizes the computing resources of a heterogeneous network remains an open problem. The authors in \cite{kiamari2017Globecom} derived a lower bound for the communication load for a CDC network where nodes have varying storage or computing capabilities. The proposed achievable scheme achieves the information-theoretical optimality of the minimum communication load for a system of $3$ nodes. The authors also demonstrated that the parameters of a heterogeneous CDC network can be translated into an optimization problem to find an efficient Map and Shuffle phase design. However, there is no optimality guarantee. In \cite{shakya2018distributed}, the authors studied CDC networks with $2$ and $3$ computing nodes where nodes have varying communication load constraints to find a lower bound on the minimum computation load. 
While \cite{kiamari2017Globecom} and \cite{shakya2018distributed} consider heterogeneous file assignments, neither work considers heterogeneous reduce function assignments. In practice, it is natural to assume that if a particular node has more computation and storage capabilities then it is advantageous to assign more reduce functions to it.

\textit{Contributions}: In this paper, we study a simplified heterogeneous computing network which consists of multiple homogeneous networks. By using the file assignment of our heterogeneous, cascaded CDC scheme\footnote{Here, ``cascaded" implies that each reduce function is redundantly computed at multiple nodes, as opposed to just one node. In this paper, we only study non-cascaded CDC.} in \cite{woolsey2019cascaded}, we demonstrate that using heterogeneous function assignments yields a simple (non-cascaded) heterogeneous CDC scheme. Our 
newly proposed CDC scheme maintains a multiplicative computation-communication load trade off such that $r-1$ nodes are simultaneously served with each transmission in the Shuffle phase. To the best of your knowledge, this is the first (non-cascaded) CDC scheme which can operate on a heterogeneous network with a large number of computing nodes. We compare the communication load of our proposed heterogeneous scheme to that of the state-of-the-art homogeneous CDC scheme \cite{li2018fundamental} with an equivalent computation load. Surprisingly, we find that if $r=\Theta(K)$, then the heterogeneous schemes outperforms the homogeneous scheme as $K$ becomes large. Finally, given the specific file and function placement of our design, we find our proposed Shuffle phase design yields a communication load that is optimal within a constant.

\paragraph*{Notation Convention}
We use $|\cdot|$ to represent the cardinality of a set or the length of a vector 
and $[n] := [1,2,\ldots,n]$. 

\section{Network Model and Problem Formulation}
\label{sec: Network Model and Problem Formulation}

The network model is similar to the network model of \cite{li2018fundamental}. We consider a distributed computing network where a set of $K$ nodes, labeled as $\{1, \ldots , K \}$, have the goal of computing $Q$ output functions and computing each function requires access to all $N$ input files. The  input files, $\{w_1 , \ldots , w_N \}$, are assumed to be of equal size of $B$ bits each. The set of $Q$ output functions is denoted by $\{ \phi_1 , \ldots \phi_Q\}$. Each node $k\in \{ 1, \ldots , K \} $ is assigned to compute a subset of output functions, denoted by $\mathcal{W}_k \subseteq \{ 1, \ldots Q \} $. Every function is assigned to exactly $1$ node. Moreover, different from \cite{li2018fundamental}, we consider heterogeneous function assignment where it is possible that $|\mathcal{W}_i|\neq |\mathcal{W}_j|$. The result of output function $i \in \{1, \ldots Q \}$ is $u_i = \phi_i \left( w_1, \ldots , w_N \right)$.

Alternatively, an output function can be computed  using ``Map" and ``Reduce" functions such that
\be
u_i = h_i \left( g_{i,1}\left( w_1 \right), \ldots , g_{i,N}\left( w_N \right) \right),
\ee
where for every output function $i$ there exists a set of $N$ Map functions $\{ g_{i,1}, \ldots , g_{i,N}\}$ and one Reduce function $h_i$. Furthermore, we define the output of the Map function, $v_{i,j}=g_{i,j}\left( w_j \right)$, as the \textit{intermediate value} resulting from performing the Map function for output function $i$ on file $w_j$. There are a total of $QN$ intermediate values  each with a size of $T$ bits.

The MapReduce distributed computing framework allows nodes to compute output functions without having access to all $N$ files. Instead, each node $k$ has access a subset of the $N$ files labeled as $\mathcal{M}_k \subseteq \{ w_1, \ldots , w_N\}$. We consider the more general heterogeneous networks where the number of files stored at each nodes varies, such that it is possible that $|\mathcal{M}_i|\neq |\mathcal{M}_j|$. Every node will compute all $Q$ intermediate values for each of its locally available files.

As every file is assigned to at least one node, collectively, the nodes use the Map functions to compute every intermediate value in the {\em Map} phase at least once. Then, in the {\em Shuffle} phase, nodes multicast the computed intermediate values among one another via a shared link. 
The Shuffle phase is necessary so that each node can receive the necessary intermediate values that it could not compute itself. Finally, in the {\em Reduce} phase, nodes use the Reduce functions with the appropriate intermediate values as inputs to compute the assigned output functions. 

\begin{table*}[t]
  \centering
  \caption{ Heterogeneous CDC Example, $K=7$, $r=3$, $Q=11$, $N=12$}
  \normalsize
  \begin{tabular}{ |>{\centering}m{1.5cm}||>{\centering}m{1.8cm}|>{\centering}m{1.8cm}||>{\centering}m{1.8cm}|>{\centering}m{1.8cm}||
  >{\centering}m{1.8cm}|>{\centering}m{1.8cm}|>{\centering}m{1.8cm}| }
 \hline
 Node & $1$ & $2$ & $3$ & $4$ & $5$& $6$ & $7$ \\

 \hline
  & $w_1$, $w_2$, & $w_7$, $w_8$,   & $w_1$, $w_3$, & $w_2$, $w_4$,   & $w_1$, $w_2$, & $w_3$, $w_4$,  & $w_5$, $w_6$, \\
  Mapped Files           &$w_3$, $w_4$,   & $w_9$, $w_{10}$,  &$w_5$, $w_7$, & $w_6$, $w_8$,  & $w_7$, $w_8$ & $w_9$, $w_{10}$  & $w_{11}$, $w_{12}$ \\
& $w_5$, $w_6$ & $w_{11}$, $w_{12}$ & $w_9$, $w_{11}$ & $w_{10}$, $w_{12}$ & & & \\

 \hline
   Assigned Functions & $1$, $2$ & $3$, $4$ & $5$, $6$ & $7$, $8$ & $9$ & $10$ & $11$ \\

 \hline
    & $v_{5,2}\oplus v_{9,3}$  &  & $v_{1,7}\oplus v_{9,5}$   & & $v_{2,7}\oplus v_{6,2}$ &                          &    \\
            & $v_{5,4}\oplus v_{10,5}$ &  & $v_{1,9}\oplus v_{10,1}$  & &                         & $v_{2,9}\oplus v_{6,4}$  &    \\
            & $v_{5,6}\oplus v_{11,1}$ &  & $v_{1,11}\oplus v_{11,3}$ & &                         &                          & $v_{2,11}\oplus v_{6,6}$   \\
            & $v_{7,1}\oplus v_{9,4}$  & & & $v_{1,8}\oplus v_{9,6}$   & $v_{2,8}\oplus v_{8,1}$ &                          &    \\
            & $v_{7,3}\oplus v_{10,6}$ & & & $v_{1,10}\oplus v_{10,2}$  &                         & $v_{2,10}\oplus v_{8,3}$  &    \\
   Shuffle  & $v_{7,5}\oplus v_{11,2}$ & & & $v_{1,12}\oplus v_{11,4}$ &                         &                          & $v_{2,12}\oplus v_{8,5}$   \\
            & & $v_{5,8}\oplus v_{9,9}$  & $v_{3,1}\oplus v_{9,11}$  & & $v_{4,1}\oplus v_{6,8}$ &                          &    \\
            & & $v_{5,10}\oplus v_{10,7}$ & $v_{3,3}\oplus v_{10,11}$ & &                         & $v_{4,3}\oplus v_{6,10}$  &    \\
            & & $v_{5,12}\oplus v_{11,7}$ & $v_{3,5}\oplus v_{11,9}$ & &                        &                          & $v_{4,5}\oplus v_{6,12}$   \\
            & & $v_{7,7}\oplus v_{9,10}$  & & $v_{3,2}\oplus v_{9,12}$  & $v_{4,2}\oplus v_{8,7}$ &                          &    \\
            & & $v_{7,9}\oplus v_{10,8}$ & & $v_{3,4}\oplus v_{10,12}$ &                         & $v_{4,4}\oplus v_{8,9}$  &    \\
            & & $v_{7,11}\oplus v_{11,8}$ & & $v_{3,6}\oplus v_{11,10}$ &                        &                          & $v_{4,6}\oplus v_{8,11}$   \\
 \hline
  \end{tabular}
  \label{table: exmp1}
\end{table*}

This distributed computing network design yields two important performance parameters: the computation load, $r$, and the communication load, $L$. The computation load is defined as the number of times each intermediate value is computed among all computing nodes. In other words, the computation load is the number of intermediate values computed in the Map phase normalized by the total number of unique intermediate values, $QN$. The communication load is defined as the amount of traffic load (in bits) among all the nodes in the Shuffle phase 
normalized by $QNT$. Moreover, we define $L^*$ as the infimum of the communication load of all achievable Shuffle phases given a particular file and function assignment.

\section{An Example}

In the following example, there are $3$ sets of nodes, $\mathcal{K}_1~=~\{ 1, 2 \}$, $\mathcal{K}_2 = \{ 3, 4 \}$ and $\mathcal{K}_3 = \{ 5, 6, 7 \}$, where each set collectively has the storage capacity to store all $N=12$ files. More specifically, each node of $\mathcal{K}_1 $ and $\mathcal{K}_2$ can store half of the files and each node of $\mathcal{K}_3$ can store one-third of the files. Each file is assigned to a set of $3$ nodes such that it contains one node from each set $\mathcal{K}_1 $, $\mathcal{K}_2 $ and $\mathcal{K}_3$. For example, file $w_1$ is assigned to the nodes of $\{1,3,5 \}$ and file $w_9$ is assigned to the nodes of $\{2,3,6 \}$. All of the files assignments are found in Table \ref{table: exmp1}. In total there are $N=12$ files. In the Map phase, the nodes will compute all intermediate values from their locally available files. Since every file is assigned to $3$ nodes, we find $r=3$.

The reduce function assignment is also shown in Table \ref{table: exmp1}. Different from previous works in CDC, nodes are assigned a varying number of reduce functions. Intuitively, we assign more reduce functions to nodes which have larger storage and computing capability. Therefore, we assign $2$ reduce functions to the nodes of $\mathcal{K}_1 $ and $\mathcal{K}_2$ and just $1$ reduce function to the nodes of $\mathcal{K}_3$. The reason we assigned this specific number of reduce functions to each node will become clear when we discuss the Shuffle phase. In total there are $Q=11$ reduce functions.

In the Shuffle phase, we consider every set of $3$ nodes such that it contains one node from each set $\mathcal{K}_1 $, $\mathcal{K}_2 $ and $\mathcal{K}_3$ (similar to the file assignment). We call each of these sets a \textit{multicast group}. Within each multicast group, nodes send coded pairs of intermediate values to the other nodes. For example, consider the node set $\{2,3,6 \}$. We are interested in intermediate values that one node requests and the other two have computed. Both nodes $2$ an $3$ have access to files $w_7$ and $w_{11}$, but node $6$ does not, therefore, node $6$ requests $v_{10,7}$ and $v_{10,11}$ from nodes $2$ and $3$. Furthermore, nodes $2$ and $6$ have access to file $w_{10}$, but node $3$ does not, therefore, node $3$ requests $v_{5,10}$ and $v_{6,10}$ from nodes $2$ and $6$; and nodes $3$ and $6$ have access to file $w_3$, but node $2$ does not, therefore, node $2$ requests $v_{3,3}$ and $v_{4,3}$. In this way, node $2$, $3$ and $6$ can transmit $v_{5,10}\oplus v_{10,7}$,  $v_{3,3}\oplus v_{10,11}$ and $v_{4,3}\oplus v_{6,10}$, respectively, among themselves. By using locally computed intermediate values to cancel ``interference", each node can recover its requests. All of the transmissions of the Shuffle phase are shown in Table \ref{table: exmp1}. Each row of transmissions represents one multicast group.

Assigning a varying number of reduce functions to the nodes has actually created symmetry among the multicast groups. Here, symmetry means each node of the group requests the same number of intermediate values from the other nodes of the group. For example, consider the node set $\{2,3,6 \}$. If every node was only assigned one reduce function the following would occur. Since there is only one file that nodes $3$ and $6$ have and node $2$ does not, node $2$ would only request one intermediate value from nodes $3$ and $6$. Similarly, node $3$ would request one intermediate values from nodes $2$ and $6$. However, there are two files that nodes $2$ and $3$ have that node $6$ does not, and therefore, node $6$ requests $2$ intermediate values. The group would be asymmetric and there is not a simple transmission policy to serve the nodes' requests. A simple design solution to create symmetry within this multicasting group is to assign two reduce functions to nodes $2$ and $3$ and just one reduce function to node $6$.

The communication load can be calculated by accounting for the $2\cdot 2 \cdot 3 = 12$ node sets of interest, where within each set, there are $3$ transmissions of size $T$ bits. By normalizing by $QNT$ we find the communication load of the coded scheme is $L_{\rm c} = \frac{36}{12 \cdot 11} = \frac{3}{11}$. We can compare this to the uncoded communication load, where each requested intermediate value is transmitted alone. To compute the uncoded communication load, we count the number of intermediate values each node requests. Since the $4$ nodes of $\mathcal{K}_1 $ and $\mathcal{K}_2$ request $6\cdot 2 = 12$ intermediate values each and the $3$ nodes of $\mathcal{K}_3$ request $8$ intermediate values each, we find $L_{\rm u} = \frac{4\cdot 12 + 3 \cdot 8}{12\cdot 11} = \frac{6}{11}$. In this case, it is clear that $L_{\rm c} = \frac{1}{2}L_{\rm u}$ since for the coded Shuffle policy every requested intermediate value is transmitted in coded pairs. In the general CDC scheme proposed here, we will see that $L_c = \frac{1}{r-1}\cdot L_{\rm u}$.

\section{General Achievable Scheme}
\label{sec: gen_sd}

In this section, we describe the general heterogeneous CDC scheme. We take advantage of many homogeneous networks and combine them into a heterogeneous one. However, we do require that the fraction of files that each node can store to be of the form $\frac{1}{m}$ where $m \in \mathbb{Z}^+$ and $m \geq 2$. Furthermore, each homogeneous computing network must be able to store file library $r'$ times for some $r' \in \mathbb{Z}^+$. The general scheme is described in more detail below.

Consider $K$ computing nodes comprised of $P$ disjoint sets of nodes, $\mathcal{C}_1, \ldots , \mathcal{C}_P$, where for all $p \in [P]$ the storage capacity of every node $k \in \mathcal{C}_p$ is $\frac{1}{m_p}NF$ bits such that $m_p,r_p\in \mathbb{Z}^+$ and $m_p \geq 2$ where we define $r_p \triangleq \frac{1}{m_p}\cdot|\mathcal{C}_p| $. Furthermore, for all $p \in [P]$, we split $\mathcal{C}_p$ into $r_p$ disjoint, equal-size subsets. In this way, the heterogeneous computing network is comprised of $r$ node sets, $\mathcal{K}_1 ,\ldots , \mathcal{K}_r$, where $r = \sum_{p=1}^{P}r_p$. Each $\mathcal{K}_i$ is a set of nodes with the same storage constraint that are collectively capable of storing the file library exactly once. More rigorously, for all $i \in [r]$, we find $\mathcal{K}_i \subseteq \mathcal{C}_p$ and $|\mathcal{K}_i|=m_p$ for some $p\in[P]$. Moreover, we find $K= \sum_{i=1}^{r}|\mathcal{K}_i|=\sum_{p=1}^{P}m_p \cdot {r_p}$. Furthermore, define
\be
X \triangleq \prod_{i=1}^{r}|\mathcal{K}_i|=\prod_{p=1}^{P}m_p^{r_p}
\ee
and $Y$ as the least common multiple (LCM) of $\{m_1-1, m_2-1, \ldots , m_p-1 \}$.

To assign the files do the following. Consider every set of nodes such that it contains exactly $1$ node from each set $\mathcal{K}_i$ for all $i \in [r]$. There are $X$ distinct sets which we label as $\mathcal{T}_1,\ldots ,\mathcal{T}_{X}$. Split the $N$ files into $X$ disjoint, equal-size sets of size $\eta_1$ files such that $N = \eta_1 X$ and $\eta_1 \in \mathbb{Z}^+$. These file sets are labeled as $\mathcal{B}_{1},\ldots,\mathcal{B}_{X}$. For all $n \in [X]$, assign the files of $\mathcal{B}_{n}$ to the nodes of $\mathcal{T}_n$. Therefore, the set of files available to node $k$ is
\be
\mathcal{M}_k=\bigcup\limits_{n : k\in \mathcal{T}_n}\mathcal{B}_n.
\ee

To assign the reduce functions do the following. We split the $Q$ functions into $K$ disjoint sets, labeled $\mathcal{W}_{1},\ldots,\mathcal{W}_{K}$, where, in general, the sets may be different sizes. We require that $Q~=~\eta_2 Y \sum_{p=1}^{P}\frac{r_p m_p}{m_p - 1}$ where $\eta_2 \in \mathbb{Z}^+$. For each node $k \in [K]$, define a reduce function set, $\mathcal{W}_{k}$, such that $|\mathcal{W}_{k}|~=~\frac{\eta_2Y }{m_p - 1}$ where $k \in \mathcal{C}_p$. The Reduce functions of $\mathcal{W}_{k}$ are assigned to node $k$.

The Map, Shuffle and reduce phases are defined as follows.

\begin{itemize}
\item {\bf Map Phase:} Each node $k\in[K]$ computes every intermediate value, $v_{i,j}$, such that $i\in [Q]$ and
$w_j\in \mathcal{M}_k$.

\item {\bf Shuffle Phase:} For all $n \in [X]$ do the following. For every node $z\in\mathcal{T}_n$, define a set of intermediate values
\be
\mathcal{V}_{\mathcal{T}_n\setminus z}^{\{z\}} = \left\{ v_{i,j} : i \in \mathcal{W}_z ,w_j \notin \mathcal{M}_z, w_j \in \bigcap\limits_{k \in \mathcal{T}_n\setminus z} \mathcal{M}_k ,  \right\}
\ee
which is the set of intermediate values requested by node $z$ for which each node of $\mathcal{T}_n\setminus z$ has computed. Furthermore, $\mathcal{V}_{\mathcal{T}_n\setminus z}^{\{z\}}$ is split into $r-1$ disjoint sets of equal size denoted by $\left\{ \mathcal{V}_{\mathcal{T}_n\setminus z}^{\{z\},\sigma_1},\ldots , \mathcal{V}_{\mathcal{T}_n\setminus z}^{\{z\},\sigma_{r-1}} \right\}=\mathcal{V}_{\mathcal{T}_n\setminus z}^{\{z\}}$ where $\{ \sigma_1,\ldots , \sigma_{r-1}\}=\mathcal{T}_n\setminus z$. Each node $k\in \mathcal{T}_n$ multicasts
    \be
    \label{eq: 1_trans_eq1}
    \bigoplus \limits_{z\in \mathcal{T}_n\setminus k} \mathcal{V}_{\mathcal{T}_n\setminus z}^{\{z\},k}
    \ee
to the nodes of $\mathcal{T}_n \setminus k$.

\item {\bf Reduce Phase:} For all $k\in [K]$, node $k$ computes all output values $u_i$ such that $i\in \mathcal{W}_k$.
\end{itemize}

\section{Achievable Computation and Communication Load}

In this section, we first derive the communication load of an uncoded Shuffle phase, $L_{\rm u}$, using the file and function assignment of Section \ref{sec: gen_sd}. Note that $L_{\rm u}$ represents the fraction of intermediate values which are requested by any node. Then, we demonstrate that the communication load using the Shuffle phase of Section \ref{sec: gen_sd} is $L_{\rm c} = \frac{1}{r-1}\cdot L_{\rm u}$. More formally, we define $L_{\rm u}$ and $L_{\rm c}$ as functions of $m_1 , \ldots , m_P$ and $r_1 , \ldots , r_P$ which defines the number of nodes and their corresponding storage constraints of the heterogeneous computing network.

\begin{theorem}
\label{theorem: 1}
Given $P$ sets of computing nodes, $\mathcal{C}_1 , \ldots , \mathcal{C}_P$, such that for all $p\in[P]$, each node $k\in\mathcal{C}_p$ has the storage capacity of $\frac{NF}{m_p}$ bits and $m_p , r_p \in \mathbb{Z}^+$ where $r_p = \frac{|\mathcal{C}_p|}{m_p}$, and using the file and function assignment defined in Section \ref{sec: gen_sd}, the uncoded communication load is
\begin{align}
L_{\rm u}(m_1 , \ldots , m_P, r_1 , \ldots , r_P) = \frac{r}{\sum_{p=1}^{P}\frac{r_p m_p}{m_p-1}}.
\end{align}
\end{theorem}

\begin{IEEEproof}
For all $p \in [P]$,  the number of files a node $k~\in~\mathcal{K}_j~\subseteq~\mathcal{C}_p$ has local access to is
\begin{align}
|\mathcal{M}_k| &= \eta_1 \prod\limits_{i\in[r]\setminus j}|\mathcal{K}_i| = \frac{\eta_1 X}{|\mathcal{K}_j|} = \frac{N}{m_p}.
\end{align}
We count the number of intermediate values that are requested by any node and normalize by $QN$
\begin{align}
L_u & (m_1 , \ldots , m_P, r_1 , \ldots , r_P) \nonumber \\
 &= \frac{1}{QN}\sum_{k\in[K]}  | \left \{ v_{i,j} : i \in \mathcal{W}_k , w_j \notin \mathcal{M}_k  \right \} | \\
&= \frac{1}{QN}\sum_{k\in[K]} \left | \mathcal{W}_k \right|\times \left( N- \left |  \mathcal{M}_k   \right| \right) \\
&= \frac{1}{QN}\sum_{p\in[P]} \sum_{k\in\mathcal{C}_p} \left | \mathcal{W}_k \right|\times \left( N- \left |  \mathcal{M}_k   \right| \right) \\
&= \frac{1}{QN}\sum_{p\in[P]} \sum_{k\in\mathcal{C}_p} \frac{\eta_2Y }{m_p - 1} \cdot \left( N- \frac{N} {m_p} \right) \\
&= \frac{1}{Q}\sum_{p\in[P]} r_p m_p \frac{\eta_2Y }{m_p - 1}\left( \frac{m_p-1}{m_p} \right) \\
&= \frac{\eta_2 Y \sum_{p\in[P]} r_p }{\eta_2 Y \sum_{p=1}^{P}\frac{r_p m_p}{m_p - 1} } = \frac{r}{\sum_{p=1}^{P}\frac{r_p m_p}{m_p-1}}
\end{align}
where $|\mathcal{C}_p| = r_p m_p$ for all $p \in [P]$.

\end{IEEEproof}

The following theorem states the communication load of the Shuffle phase which uses coded communication.

\begin{theorem}
\label{theorem: 2}
Given $P$ sets of computing nodes, $\mathcal{C}_1 , \ldots , \mathcal{C}_P$, such that for all $p\in[P]$, each node $k\in\mathcal{C}_p$ has the storage capacity of $\frac{NF}{m_p}$ bits and $m_p , r_p \in \mathbb{Z}^+$ where $r_p = \frac{|\mathcal{C}_p|}{m_p}$, and using the file and function assignment defined in Section \ref{sec: gen_sd}, the coded communication load which uses the Shuffle phase of Section \ref{sec: gen_sd} is
\begin{align}
L_{\rm c} (m_1 , \ldots , m_P, & r_1 , \ldots , r_P) \nonumber \\
&= \frac{1}{r-1}\cdot\frac{r}{\sum_{p=1}^{P}\frac{r_p m_p}{m_p-1}} \\
&= \frac{1}{r-1} \cdot L_{\rm u}(m_1 , \ldots , m_P, r_1 , \ldots , r_P).
\end{align}
\end{theorem}

\begin{IEEEproof}
For any $n \in [X]$, and for all $z \in \mathcal{T}_n$ where $z \in \mathcal{K}_p$, we find
\begin{align}
\big| & \mathcal{V}_{\mathcal{T}_n\setminus z}^{\{z\}} \big| \nonumber \\
&= \left | \mathcal{W}_z \right | \times \left | \left \{ w_j: w_j \notin \mathcal{M}_z, w_j \in \bigcap\limits_{k \in \mathcal{T}_n\setminus z} \mathcal{M}_k , \right \} \right| \\
&= \left | \mathcal{W}_z \right | \cdot \eta_1 \left | \left \{ \mathcal{T}_{n'} : \{\mathcal{T}_n\setminus z \} \subset \mathcal{T}_{n'}, z \notin \mathcal{T}_{n'}, n' \in [X] \right \} \right | \\
&= \left | \mathcal{W}_z \right | \cdot \eta_1 \left | \left \{ \mathcal{T}_{n'} : \{\mathcal{T}_n\setminus z \} \cup k = \mathcal{T}_{n'}, k \in \mathcal{K}_p\setminus z,  \right \} \right | \\
&= \left | \mathcal{W}_z \right | \cdot \eta_1 \left | \left \{ k :  k \in \mathcal{K}_p\setminus z,  \right \} \right | \\
&= \left | \mathcal{W}_z \right | \cdot \eta_1 (|\mathcal{K}_p|-1) \\
&= \frac{\eta_2Y }{m_p - 1} \cdot \eta_1 (m_p - 1)\\
&= \eta_1 \eta_2 Y.
\end{align}

We consider $X$ node groups of size $r$ nodes, where for each group, every node of that group transmits a coded message of size $\big|  \mathcal{V}_{\mathcal{T}_n\setminus z}^{\{z\}} \big| / (r-1)$, therefore, the communication load is
\begin{align}
L_{\rm c} ( m_1 , \ldots & , m_P, r_1 , \ldots , r_P) \nonumber \\
&= \frac{1}{QN}\cdot X \cdot r \cdot  \frac{\big|  \mathcal{V}_{\mathcal{T}_n\setminus z}^{\{z\}} \big|}{r-1} \\
&= \frac{1}{\left[\eta_2 Y \sum_{p=1}^{P}\frac{r_p m_p}{m_p - 1}\right]\eta_1 X}\cdot X \cdot r \cdot \frac{\eta_1 \eta_2 Y}{r-1} \\
&= \frac{1}{r-1}\cdot\frac{r}{\sum_{p=1}^{P}\frac{r_p m_p}{m_p-1}}.
\end{align}

\end{IEEEproof}

The communication load $L_{\rm c}$, is comprised of two parts: the local computing gain, $L_{\rm u}$, and the global computing gain, $\frac{1}{r-1}$. The local computing gain represents the normalized number of intermediate values that must be shuffled. As nodes have access to a larger fraction of the files, the nodes will inherently request less in the Shuffle phase. The global computing gain stems from the fact that with the coded design every transmission serves $r-1$ nodes with distinct requests.

\section{Comparison to State-of-the-Art Homogeneous CDC Achievable Scheme}

\label{sec: disc_sg1}
For a given computation load, $r$, and number of nodes, $K$, the communication load of the achievable homogeneous CDC scheme of \cite{li2018fundamental} is
\be
L_1 = \frac{1}{r}\cdot \left( 1 - \frac{r}{K} \right).
\ee
This can be broken up into the local computing gain, $1-\frac{r}{K}$, and the global computing gain, $\frac{1}{r}$. In the following we show that the local computing gain of our new heterogeneous design can be less than the local computing gain of $L_1$.

Since $\sum_{p=1}^{P} \frac{r_p}{r} = 1$ and $\frac{m_p}{m_p-1}$ is a convex function of $m_p$ for $m_p > 1$, by Jensen's inequality

\begin{align}
\frac{\sum_{p=1}^{P}\frac{r_p m_p}{m_p-1}}{r} &= \sum_{p=1}^{P} \frac{r_p}{r}\cdot \frac{m_p}{m_p-1} \\
&\geq \frac{\sum_{p=1}^{P} \frac{r_p m_p}{r}}{\left[\sum_{p=1}^{P} \frac{r_p m_p}{r}\right] - 1} \\
&= \frac{\frac{K}{r}}{\frac{K}{r}-1} \\
&= \frac{K}{K-r}
\end{align}
where $\sum_{p=1}^{P}r_p m_p = \sum_{p=1}^{P}|\mathcal{C}_p| = K$. In other words,

\be
L_{\rm u} \leq \frac{K-r}{K} = 1-\frac{r}{K}
\ee
and
\be
L_{\rm c} \leq \frac{1}{r-1}\cdot \left( 1-\frac{r}{K} \right).
\ee

The local computing gain for our heterogeneous design is upper bounded by the local computing gain of the homogeneous CDC scheme of \cite{li2018fundamental}. For this reason $L_{\rm c}$ can be less than $L_1$ for a given $r$ and $K$. For example, given a heterogeneous network defined by $m_1 = 2$, $r_1 = 4$ and $m_2 = 8$, $r_2 = 2$ we find $r=6$, $K=24$, and using the new heterogeneous design $L_{\rm u} = \frac{7}{12} \approx 0.583$ and $L_{\rm c} = \frac{7}{ 60}\approx 0.117$. However, for an equivalent homogeneous network with $r=6$ and $K=24$ the local computing gain is $1-\frac{r}{K} = \frac{3}{4} = 0.75$ and communication load using the coded design is $L_1 = \frac{1}{8} = 0.125$.

\begin{remark}
  In \cite{li2018fundamental}, $L_1$ was proven to be a lower bound on the communication load for a given $r$ and $K$, however, the proof uses the assumption that every node is assigned the same number of reduce functions. If the reduce functions can be assigned in a heterogeneous fashion, the communication load lower bound derived in \cite{li2018fundamental} does not apply.
\end{remark}

To better understand the comparison of the schemes for a large number of computing nodes, $K$, we show the following. Consider the case where $r$, and $r_1,\ldots , r_P$ for the heterogeneous case, are fixed as $K$ becomes large. In other words, the fraction of files each node has access to decreases as $K$ grows (i.e. for all $p\in[P]$, $m_p \rightarrow \infty$ as $ K \rightarrow \infty$). In this case
\be
\lim_{K\rightarrow \infty} \frac{L_{\rm c}}{L_1} = \frac{r}{r-1}.
\ee
In other words, $\frac{L_{\rm c}}{L_1} = \Theta (1)$.

Alternatively, we can observe the case where $r$, and $r_1,\ldots , r_P$ for the heterogeneous case, grow linearly with $K$ and the fraction of files available to each node is constant. $m_1 , \ldots , m_p$ are constant for the heterogeneous scheme. In this case
\be
\lim_{K\rightarrow \infty} \frac{L_{\rm c}}{L_1} = \lim_{K\rightarrow \infty} \frac{r}{r-1} \cdot \frac{L_{\rm u}}{1-\frac{r}{K}} = \frac{L_{\rm u}}{1-\frac{r}{K}}  \leq 1
\ee
where, $L_{\rm u}$ is a constant since $\frac{r_1}{r}, \ldots , \frac{r_P}{r}$  are constants, and $1-\frac{r}{K}$ is constant since $\frac{r}{K}$ is constant. Again, we see that $\frac{L_{\rm c}}{L_1} = \Theta (1)$.

\section{Optimality}
As shown in the previous section, the lower bound of the communication load derived in \cite{li2018fundamental} does not apply when reduce functions are heterogeneously assigned to the computing nodes.  In the following we discuss communication load bounds for two scenarios. First, we demonstrate a lower bound on communication load when considering all possible file and function assignments for a given $r$ and $K$. Next, we provide a lower bound on the communication load when we use the specific file and function assignment of the heterogeneous design in Section \ref{sec: gen_sd} is used.

A trivial bound on the communication load is $L \geq 0$. Given $r$ and $K$, the following file and function assignment and Shuffle phase design will yield a communication load meeting this bound. Pick $r$ nodes and assign the entire file library to each of the nodes. Furthermore, for each function, assign it to one of the $r$ nodes with access to the entire file library. As every node is able to compute all the necessary intermediate values itself, no Shuffle phase is required and $L=0$. Note that, in this context, we do not consider any storage or computing limitations on the nodes, rather, we show that optimizing the communication load over all possible function and file assignments is not an interesting problem.

The question remains as to the optimality of the proposed Shuffle phase of Section \ref{sec: gen_sd}. Based on the approach introduced in \cite{wan2016caching,wan2016optimality} for coded caching, we derive the following theorem which provides a lower bound on the entropy of all transmissions in the Shuffle phase given a specific function and file placement and a permutation of the computing nodes.

\begin{theorem}
\label{theorem: bound}
  Given a set of $K$ nodes, labeled as $\mathcal{K}$, in order for every node $k\in\mathcal{K}$ to have access to all intermediate values necessary to compute functions of $\mathcal{W}_k$, the entropy of the collective transmissions by all nodes, $H(X_\mathcal{K})$, is bounded by
  \be \label{eq: bound_eq1}
  H(X_\mathcal{K}) \geq \sum_{i=1}^{K} H\left(V_{\mathcal{W}_{k_i},:}|V_{:,\mathcal{M}_{k_i}},Y_{\{k_1,\ldots, k_{i-1} \}}\right)
  \ee
  where $k_1, \ldots , k_K$ is some permutation of $[K]$, $V_{\mathcal{W}_{k_i},:}$ is the set of intermediate values necessary to compute the functions of $\mathcal{W}_{k_i}$, $V_{:,\mathcal{M}_{k_i}}$ is set of intermediate values which can be computed from the file set $\mathcal{M}_{k_i}$ and $Y_{\{k_1,\ldots, k_{i-1} \}}$  is the union of the set of intermediate values necessary to compute the functions of  $\bigcup_{j=1}^{i-1}\mathcal{W}_{k_j}$ and the set of intermediate values which can be computed from files of $\bigcup_{j=1}^{i-1}\mathcal{M}_{k_j}$.
\end{theorem}

Theorem \ref{theorem: bound} is proved in Appendix \ref{appendix: bound}. In the next theorem, we demonstrate that given the specific function and file placement of Section \ref{sec: gen_sd}, the Shuffle phase design of Section \ref{sec: gen_sd} yields a communication load that is within a constant of the lower bound.

\begin{theorem}
\label{thm: optimality}
For a computing network of $K$ nodes with the file assignments, $\mathcal{M}_1 , \ldots \mathcal{M}_K$, and function assignments, $\mathcal{W}_1 , \ldots \mathcal{W}_K$ as defined in Section \ref{sec: gen_sd}, define $L^*$ to be the infimum of the communication load over all possible Shuffle phases, then
\be
 L_{\rm c} \leq 4L^*
\ee
where $L_{\rm c}$ is the communication load from the coded Shuffle phase design of Section \ref{sec: gen_sd}.
\end{theorem}

In Appendix \ref{sec: opt_pf} we show how Theorem \ref{theorem: bound} can be used to prove Theorem \ref{thm: optimality}.

\section{Conclusion}

In this work, we have introduced a novel approach to the design of CDC networks where reduce functions are assigned in a heterogeneous fashion. Moreover, the achievable scheme presented here maintains a multiplicative computation-communication trade-off similar to that of the homogeneous CDC network designs. Surprisingly, the optimal trade-off derived in \cite{li2018fundamental} no longer applies when functions are heterogeneously assigned and the communication load of a heterogeneous network can be less than that of an equivalent homogeneous CDC network. Given our proposed heterogeneous file and function assignment, we derived a lower bound of the communication load and demonstrated our Shuffle phase yields a communication load that is optimal within a constant factor. It will be interesting to find other achievable schemes with heterogeneous function assignments and a more general communication load bound given a set of set of storage capacity requirements of the computing nodes.

\appendices

\section{Proof of Theorem \ref{theorem: bound}}
\label{appendix: bound}
In this proof, we use the following notation: $\mathcal{K}$ is the set of all nodes, $X_{\mathcal{K}}$ represents the collection of all transmissions by all nodes in $\mathcal{K}$, $\mathcal{W}_{\mathcal{S}}$ is the set of functions assigned to at least on node of $\mathcal{S}$, $\mathcal{M}_{\mathcal{S}}$ is the set files locally available to at least one node in $\mathcal{S}$, $V_{\mathcal{W}_{\mathcal{S}_1},\mathcal{M}_{\mathcal{S}_2}}$ is the set of intermediate values needed to compute the functions of $\mathcal{W}_{\mathcal{S}_1}$ and computed from the files of $\mathcal{M}_{\mathcal{S}_2}$. Finally, we define the following
\be
Y_\mathcal{S} \triangleq \left(V_{\mathcal{W}_\mathcal{S},:},V_{:,\mathcal{M}_\mathcal{S}}\right)
\ee
where ``$:$" is used to denote all possible indices.

%
Given all the transmissions from all nodes, $X_\mathcal{K}$, and intermediate values which can be locally computed by a node $k$, $V_{:,\mathcal{M}_k}$, node $k$ needs to have access to all intermediate values necessary for its assigned functions, $V_{\mathcal{W}_k,:}$, therefore
\be
H(V_{\mathcal{W}_k,:} | X_\mathcal{K}, V_{:,\mathcal{M}_k}) = 0.
\ee

Given this assumption, it is clear that
\begin{align}
H(X_\mathcal{K}) &\geq H(X_\mathcal{K}|V_{:,M_{k_1}}) \nonumber\\
&= H(X_\mathcal{K},V_{\mathcal{W}_{k_1},:}|V_{:,M_{k_1}}) - H(V_{\mathcal{W}_{k_1},:} | X_\mathcal{K}, V_{:,\mathcal{M}_{k_1}}) \nonumber\\
&= H(X_\mathcal{K},V_{\mathcal{W}_{k_1},:}|V_{:,M_{k_1}}) \nonumber\\
&= H(V_{\mathcal{W}_{k_1},:}|V_{:,M_{k_1}}) + H(X_\mathcal{K}|V_{\mathcal{W}_{k_1},:},V_{:,M_{k_1}})\nonumber\\
&=H(V_{\mathcal{W}_{k_1},:}|V_{:,M_{k_1}}) + H(X_\mathcal{K}|Y_{k_1}). \label{eq: claim 1 2}
\end{align}
Similarly,
\begin{align}
H&(X_\mathcal{K}|Y_{\{k_1,\ldots k_{i-1}\}}) \nonumber\\
&\geq H(X_\mathcal{K}|V_{:,M_{k_i}},Y_{\{k_1,\ldots k_{i-1}\}}) \nonumber\\
&= H(X_\mathcal{K},V_{\mathcal{W}_{k_i},:}|V_{:,M_{k_i}},Y_{\{k_1,\ldots k_{i-1}\}}) \nonumber\\
&\text{  }\text{ }\text{ }- H(V_{\mathcal{W}_{k_i},:} | X_\mathcal{K}, V_{:,\mathcal{M}_{k_i}},Y_{\{k_1,\ldots k_{i-1}\}}) \nonumber\\
&= H(X_\mathcal{K},V_{\mathcal{W}_{k_i},:}|V_{:,M_{k_i}},Y_{\{k_1,\ldots k_{i-1}\}}) \nonumber\\
&= H(V_{\mathcal{W}_{k_i},:}|V_{:,M_{k_i}},Y_{\{k_1,\ldots k_{i-1}\}}) \nonumber \\
&\text{ }\text{ }\text{ }+ H(X_\mathcal{K}|V_{\mathcal{W}_{k_i},:},V_{:,M_{k_i}},Y_{\{k_1,\ldots k_{i-1}\}})\nonumber\\
&= H(V_{\mathcal{W}_{k_i},:}|V_{:,M_{k_i}},Y_{k_1,\ldots k_{i-1}}) + H(X_\mathcal{K}|Y_{\{k_1,\ldots k_i\}}). \label{eq: claim 1 3}
\end{align}
Also, since nodes can only transmit intermediate values from locally available files, we see that
\be
H(X_\mathcal{K}|Y_{\{k_1,\ldots k_K\}}) = 0.
\ee
By starting with (\ref{eq: claim 1 2}) and iteratively using the relationship of (\ref{eq: claim 1 3}) to account for all $k_i \in \mathcal{K}$, we obtain (\ref{eq: bound_eq1}) and prove Theorem \ref{theorem: bound}.

\section{Proof of Theorem \ref{thm: optimality}}
\label{sec: opt_pf}

We define a permutation of the $K$ nodes, $(k_1, \ldots , k_K)$, such that $\{ k_1 , \ldots , k_{m_p} \} = \mathcal{K}_i \subseteq \mathcal{C}_p$ for some $i\in[r]$ and $p\in[P]$ as defined in Section \ref{sec: gen_sd}. 
For $1 \leq j \leq m_p$, given all intermediate values collectively computed by nodes $k_1 , \ldots , k_j$ and all intermediate values needed by nodes $k_1 , \ldots , k_{j-1}$ to compute their respective reduce functions, the entropy of the requested intermediate values of the node $k_j$ is
\begin{align}
H &\left( \mathcal{V}_{ \mathcal{W}_{k_j} , :} | \mathcal{V}_{ : , \mathcal{M}_{k_1} } , Y_{ \{ k_1,\ldots k_{j-1} \} }\right) \nonumber \\
&= H \left( \mathcal{V}_{ \mathcal{W}_{k_j} , :} | \mathcal{V}_{ : , \mathcal{M}_{\{ k_1 , \ldots , k_{j-1} \}} } \right) \\
&= T |\mathcal{W}_{k_j}|\left( N - \bigcup\limits_{j' \in [j]} \mathcal{M}_{k_{j'}} \right) \\
&= T\cdot\frac{\eta_2Y}{m_p-1}\left( N - \sum_{j' \in [j]}|\mathcal{M}_{k_{j'}}| \right) \\
&=\frac{T\eta_2Y}{m_p-1}\left( N - \frac{jN}{m_p} \right) \\
&= \frac{T\eta_2YN}{(m_p-1)m_p}\left( m_p - j \right).
\end{align}
Furthermore, since the nodes $k_1 , \ldots , k_{m_p}$ collectively have access to all the $N$ files and compute all $QN$ intermediate values, we see that for $m_p \leq j \leq K$
\be
H \left( \mathcal{V}_{ \mathcal{W}_{k_j} , :} | \mathcal{V}_{ : , \mathcal{M}_{k_1} } , Y_{ \{ k_1,\ldots k_{j-1} \} }\right) = 0.
\ee
By using of the bound of Theorem \ref{theorem: bound}
\begin{align}
H(X_\mathcal{K}) &\geq \sum_{j=1}^{m_p-1} H \left( \mathcal{V}_{ \mathcal{W}_{k_j} , :} | \mathcal{V}_{ : , \mathcal{M}_{k_1} } , Y_{ \{ k_1,\ldots k_{j-1} \} }\right) \\
&=\sum_{j=1}^{m_p-1}\frac{T\eta_2YN}{(m_p-1)m_p}\left( m_p - j \right) \\
&= \frac{\eta_2TYN}{(m_p-1)m_p}\sum_{j=1}^{m_p-1} j \\
&= \frac{\eta_2TYN}{(m_p-1)m_p}\cdot \frac{m_p(m_p-1)}{2} \\
&= \frac{\eta_2TYN}{2}.
\end{align}
Therefore, a lower bound on the communication load is
\be
L^* \geq \frac{1}{QN} \cdot \frac{\eta_2TYN}{2} = \frac{1}{2\sum_{p=1}^{P}\frac{r_p m_p}{m_p-1}}.
\ee
Finally, we see that
\be
\frac{L_{\rm c}}{L^*} = \frac{2r}{r-1} \leq 4
\ee
for $r\geq 2$. This completes the proof of Theorem \ref{thm: optimality}.

\bibliographystyle{IEEEbib}
\bibliography{references_d2d}

\begin{thebibliography}{1}

\bibitem{li2018fundamental}
S.~Li, M.~A. Maddah-Ali, Q.~Yu, and A.~S. Avestimehr,
\newblock ``A fundamental tradeoff between computation and communication in
  distributed computing,''
\newblock {\em IEEE Transactions on Information Theory}, vol. 64, no. 1, pp.
  109--128, 2018.

\bibitem{dean2008mapreduce}
J.~Dean and S.~Ghemawat,
\newblock ``Mapreduce: simplified data processing on large clusters,''
\newblock {\em Communications of the ACM}, vol. 51, no. 1, pp. 107--113, 2008.

\bibitem{konstantinidis2018leveraging}
K.~Konstantinidis and A.~Ramamoorthy,
\newblock ``Leveraging coding techniques for speeding up distributed
  computing,''
\newblock {\em arXiv:1802.03049}, 2018.

\bibitem{woolsey2018new}
N.~Woolsey, R.~Chen, and M.~Ji,
\newblock ``A new combinatorial design of coded distributed computing,''
\newblock in {\em 2018 IEEE International Symposium on Information Theory
  (ISIT)}. IEEE, 2018, pp. 726--730.

\bibitem{kiamari2017Globecom}
M.~Kiamari, C.~Wang, and A.~S. Avestimehr,
\newblock ``On heterogeneous coded distributed computing,''
\newblock in {\em GLOBECOM 2017-2017 IEEE Global Communications Conference}.
  IEEE, 2017, pp. 1--7.

\bibitem{shakya2018distributed}
N.~Shakya, F.~Li, and J.~Chen,
\newblock ``Distributed computing with heterogeneous communication constraints:
  The worst-case computation load and proof by contradiction,''
\newblock {\em arXiv:1802.00413}, 2018.

\bibitem{woolsey2019cascaded}
N.~Woolsey, R.~Chen, and M.~Ji,
\newblock ``Cascaded coded distributed computing on heterogeneous networks,''
\newblock {\em arXiv preprint arXiv:1901.07670}, 2019.

\bibitem{wan2016caching}
K.~Wan, D.~Tuninetti, and P.~Piantanida,
\newblock ``On caching with more users than files,''
\newblock in {\em 2016 IEEE International Symposium on Information Theory
  (ISIT)}, July 2016, pp. 135--139.

\bibitem{wan2016optimality}
K.~Wan, D.~Tuninetti, and P.~Piantanida,
\newblock ``On the optimality of uncoded cache placement,''
\newblock in {\em 2016 IEEE Information Theory Workshop (ITW)}, Sept 2016, pp.
  161--165.

\end{thebibliography}

\end{document}